\documentclass[fleqn,10pt]{wlscirep}
\usepackage{amsmath, bm}
\usepackage[utf8]{inputenc}
\usepackage[T1]{fontenc}

\usepackage{lineno}
\usepackage[nodisplayskipstretch]{setspace}
\usepackage[format=plain, font={small, singlespacing}, labelfont=bf]    {caption}
\usepackage{capt-of}
\usepackage[CaptionAfterwards]{fitpage}
\usepackage{caption}
\usepackage{xr}
\usepackage{xcolor}
\usepackage{multirow}
\usepackage{colortbl}
\usepackage[all,color]{xy}
\usepackage{algpseudocode}
\usepackage{algorithm}
\usepackage{wrapfig}
\usepackage{makecell}

\PassOptionsToPackage{hyphens}{url}\usepackage{hyperref}
\allowdisplaybreaks[1]

    \usepackage{soul}
    \usepackage{xspace}

    \title{Comparing large language models and human programmers for generating programming code}
    \author[1,$\dagger$]{Wenpin Hou}
    \author[2,$\dagger$]{Zhicheng Ji}
    \affil[1]{Department of Biostatistics, The Mailman School of Public Health, Columbia University, New York City, NY, USA}
    \affil[2]{Department of Biostatistics and Bioinformatics, Duke University School of Medicine, Durham, NC, USA.}    
     \affil[$\dagger$]{Corresponding author. E-mail: wh2526@cumc.columbia.edu; zhicheng.ji@duke.edu}
    
    \begin{abstract}

We systematically evaluated the performance of seven large language models in generating programming code using various prompt strategies, programming languages, and task difficulties. GPT-4 substantially outperforms other large language models, including Gemini Ultra and Claude 2. The coding performance of GPT-4 varies considerably with different prompt strategies. In most LeetCode and GeeksforGeeks coding contests evaluated in this study, GPT-4 employing the optimal prompt strategy outperforms 85 percent of human participants. Additionally, GPT-4 demonstrates strong capabilities in translating code between different programming languages and in learning from past errors. The computational efficiency of the code generated by GPT-4 is comparable to that of human programmers. These results suggest that GPT-4 has the potential to serve as a reliable assistant in programming code generation and software development.
\end{abstract}
    
\begin{document}
    \flushbottom
    \maketitle



\section*{Introduction}

The emergence and growth of large language models (LLMs), such as GPT-4, Gemini, Claude, and Llama, are poised to revolutionize various sectors, including computer programming\cite{bubeck2023sparks, li2023can, bordt2023chatgpt, cheng2023gpt, singla2023evaluating, tarassow2023potential, cipriano2023gpt, tian2023chatgpt}, education\cite{phung2023generative, kung2023performance}, and biomedical research\cite{hou2023geneturing, hou2023celltype, hou2023vision, Shue2023-ee, lecler2023revolutionizing}. A notable feature of LLMs is their usability. They can be guided to accomplish various complex tasks through prompts, which are tailored natural language text directives specifying user expectations. This feature significantly increases the inclusivity in various domains, especially in computer programming, where programmers typically need comprehensive training in computer science and sufficient programming experience. LLMs have the potential to democratize the process of computer programming, thereby increasing the volume and diversity of the programming workforce.


The coding capabilities of LLMs have been benchmarked through two types of studies. The first type designed new benchmark datasets, including HumanEval\cite{chen2021evaluating}, MBPP\cite{austin2021program}, APPS\cite{hendrycks2021measuring}, CoNaLA\cite{yin2018learning}, and CodeContests\cite{li2022competition}, to evaluate the performance of LLMs. The second type of study assessed the performance of LLMs using programming tasks directly obtained from coding practice websites, such as LeetCode \cite{bubeck2023sparks, gpt4techreport, bordt2023chatgpt, tian2023chatgpt}. According to the GPT-4 technical report by OpenAI \cite{gpt4techreport}, GPT-4 achieved accuracies of 75.6\%, 26.3\%, and 6.7\% on easy, medium, and hard LeetCode coding tasks, respectively. A more recent study \cite{bubeck2023sparks} reports that GPT-4 reached one-attempt accuracies of 68.2\%, 40\%, and 10.7\%, and five-attempt accuracies of 86.2\%, 60.0\%, and 14.3\% on easy, medium, and hard tasks, respectively. In comparison, GPT-3.5 achieved significantly lower accuracies of 58\%, 18\%, and 1\% on easy, medium, and hard tasks, respectively \cite{tian2023chatgpt}. 

Several major limitations are evident in these studies. First, existing research primarily relies on a simplistic prompt strategy that involves repeatedly presenting the same programming task to LLMs, without exploring the potential of alternative prompt strategies. Previous studies have demonstrated that varying prompt strategies can significantly impact LLM performance \cite{wei2022chain, yao2023tree}, suggesting that LLM programming capabilities could be further enhanced through prompt engineering. Second, there is a lack of rigorous comparison between the programming abilities of LLMs and humans. Such comparisons are crucial for understanding LLM limitations and identifying ways they can effectively collaborate with humans in development environments. While one study \cite{bubeck2023sparks} compares GPT-4's performance to the acceptance rate of LeetCode programming tasks by humans, it notes that this rate includes all historical submissions, many of which may be copies of published solutions. Hence, the acceptance rate may not accurately represent the actual coding abilities of human programmers, leading to biased comparison results. Third, the focus of most studies has been exclusively on the Python programming language, leaving the performance of LLMs in other widely used languages, such as Java, Javascript, and C++, underexplored. Additionally, LLMs' capability to translate across different programming languages remains poorly understood. Fourth, prior research has predominantly concentrated on the accuracy of code generated by LLMs, neglecting other important aspects such as code execution time and the ability of LLMs to learn from coding errors. These aspects are vital for the development of programming code and software.

In this study, we conducted a systematic assessment of the ability of seven LLMs to generate programming code. We addressed the limitations identified in previous research by investigating how different prompt strategies affect LLMs' coding performance, directly comparing the coding abilities of LLMs with those of human programmers, evaluating LLMs' performance in generating and translating code across various programming languages, and examining LLMs' computational efficiency and their ability to learn from past errors. We discovered that GPT-4 is the most effective LLM in generating programming code, and that employing an optimal prompt strategy significantly enhances GPT-4's coding abilities compared to the standard prompt strategy used in previous studies. With this optimal strategy, GPT-4 outperforms 85\% of human participants in most contests on LeetCode and GeeksforGeeks, demonstrating its strong potential as a tool in programming code generation and software development.

\section*{Results}

\subsection*{Evaluation datasets}

In this study, we evaluated LLMs using programming tasks sourced from LeetCode and GeeksforGeeks, the two most visited platforms for practicing programming (Supplementary Figure 1). We selected LeetCode as the primary source for our evaluation dataset, as its programming tasks can be directly compiled by copying and pasting contents from the LeetCode website. In contrast, GeeksforGeeks does not permit copy-and-paste operations, requiring manual transcription of the programming tasks. Consequently, only a small number of tasks were compiled from GeeksforGeeks. We did not include benchmark datasets such as HumanEval and MBPP, since the performance of human programmers on these datasets is unknown.

We manually compiled 100 programming tasks from 25 randomly selected LeetCode contests. Four of these contests were held before September 2021, the knowledge cutoff of GPT-4 and GPT-3.5. The remaining contests, representing various months and years, were held after September 2021 (Supplementary Figure 2). Each LeetCode contest comprises four programming tasks with easy, medium, or hard difficulties, as determined by LeetCode. A typical LeetCode task includes a problem description, example test cases with expected outputs and explanations, constraints, and a code template (Supplementary Figure 3a-d). LeetCode assesses the correctness of a solution by checking whether it generates accurate outputs for all test cases. If a solution fails to pass all test cases, LeetCode provides error messages (Supplementary Figure 3e). LeetCode accepts solutions in various programming languages. In this study, we evaluated GPT-4's performance in generating code in Python3, C++, Java, and Javascript. For successful solutions, LeetCode reports the percentiles of memory usage and running time compared to all human submissions (Supplementary Figure 3f).

We also manually compiled 15 programming tasks from five GeeksforGeeks contests, all of which were published after September 2021. The tasks from GeeksforGeeks have content similar to that of LeetCode tasks. Additionally, GeeksforGeeks provides error messages when a solution fails to pass all the test cases. Due to the limited number of programming tasks from GeeksforGeeks collected in this study, they were used solely for comparing contest performances between LLMs and humans.

\subsection*{Prompt strategies}

We designed six prompt strategies to test the performance of LLMs in solving these programming tasks (Figure 1, Methods). In the first strategy (repeated prompt), the full programming task along with the example test case were repeatedly given to the LLM. Error messages, if any, are ignored throughout the process. This strategy is commonly used in most existing studies. The second strategy (repeated prompt without example) is identical to the first, except that it omits the example test cases. This approach tests whether an LLM benefits from the strategy of including reasoning steps in the prompt akin to the 'chain-of-thoughts' concept \cite{wei2022chain}. In the third strategy (multiway prompt), an LLM is given the full programming task and is asked to generate five different solutions simultaneously. In the fourth strategy (feedback prompt), an LLM is given the full programming task, and its output is evaluated by LeetCode. If the solution fails, the LLM receives the error message from the previous attempt as input. The fifth strategy (feedback CI prompt) is specific to GPT-4. It mirrors the fourth, but specifically requires GPT-4 to evaluate the example test cases using the code interpreter (CI). CI provides GPT-4 with a Python working environment to execute and test its code against the example test cases. In subsequent attempts, GPT-4 evaluates the entire pool of test cases, including any new failed test cases reported in LeetCode's error messages, using CI. In the sixth strategy (Python3 translation), which is also specific to GPT-4, GPT-4 translates Python3 code, generated by the feedback CI prompt strategy, into Java, JavaScript, and C++.


\subsection*{Comparing success rates of GPT-4 with different prompt strategies}

We first tested the performance of GPT-4, an LLM developed by OpenAI, with six prompt strategies on the 100 LeetCode programming tasks. Each prompt strategy had a maximum of five opportunities to solve each task. We recorded the success rates of each prompt strategy, both for solving the tasks on the first attempt (one-attempt) and within five attempts (five-attempt). 

We compared GPT-4's five-attempt success rate for tasks published before or after its knowledge cutoff. GPT-4 successfully solves all programming tasks published before September 2021 within five attempts, whereas the success rate significantly drops for tasks published after September 2021 (Supplementary Figure 4). A potential reason is that tasks before September 2021 were included as part of GPT-4's training data. To more rigorously assess GPT-4's coding performance and to ensure a more fair comparison with other LLMs, our subsequent analyses exclusively focus on contests and tasks that were published after September 2021.

Figure 2 shows the one-attempt and five-attempt success rates for different prompt strategies, programming languages, and tasks with varying difficulty levels for GPT-4. The five-attempt accuracy of the repeated prompt strategy with example cases is 86\%, 58\%, and 15\% for easy, medium, and hard LeetCode tasks, respectively, which closely resemble the accuracy of the previous study \cite{bubeck2023sparks} and demonstrate the reproducibility of results across benchmark studies. For Python3, we found that the feedback CI prompt has the best overall performance, followed by the feedback prompt. Both strategies significantly outperform the multiway prompt and repeated prompt. Specifically, the feedback CI prompt increases the five-attempt performance of the repeated prompt by 16\% for easy tasks, 48\% for medium tasks, and 120\% for hard tasks. These results suggest that different prompt strategies significantly impact GPT-4's coding ability, and choosing the optimal prompt strategy can substantially improve performance. With the feedback CI prompt, GPT-4 is able to correctly solve all easy tasks and 86\% of medium difficulty tasks with five attempts, showing that GPT-4 can reliably generate code for tasks with moderate difficulty. In addition, the repeated prompt with example cases outperforms the repeated prompt without example cases, showing that GPT-4 benefits from the reasoning provided by the example test cases in a way similar to chain-of-thoughts \cite{wei2022chain}. Moreover, GPT-4 shows comparable performance across different programming languages when using the same feedback prompt strategy. The variation in performance due to different programming languages is smaller than that caused by different prompt strategies.

\subsection*{Comparing success rates of different LLMs}

After identifying the impact of prompt strategy on the performance of LLMs, we comp ared the success rates of seven LLMs (Methods). These included Gemini Ultra and Gemini Pro, developed by Google; Claude 2, developed by Anthropic; GPT-3.5, developed by OpenAI; and Llama 2 and Code Llama, developed by Meta. For Gemini Ultra, Gemini Pro, Claude 2, GPT-3.5, and Llama 2, we utilized the feedback prompt strategy, as these LLMs lack access to environments similar to GPT-4's CI system. For Code Llama, we employed a repeated prompt strategy because it often failed to comprehend error messages and produced irrelevant results.

Similar to GPT-4, GPT-3.5 successfully solves all programming tasks before September 2021, but its success rate drops considerably afterwards. A significant performance drop is not observed for other LLMs, as shown in Supplementary Figure 4. It must be noted that LLMs other than GPT-4 and GPT-3.5 do not explicitly report their knowledge cutoffs, and their training data likely includes programming tasks collected at any time. Additionally, LLMs such as Gemini Ultra and Gemini Pro have direct access to the Internet. Thus, the performance of these LLMs reported in this study may be exaggerated compared to their performance on programming tasks whose solutions cannot be found in the training data or online. These results should be interpreted with caution.

Figure 2 compares the performance of various LLMs to GPT-4, employing different prompt strategies. Despite the potential advantages of other LLMs discussed previously, GPT-4 consistently outperforms all other LLMs, even when utilizing the same prompt strategy. For instance, the five-attempt success rate for medium tasks by Gemini Ultra, the best-performing LLM aside from GPT-4, is 36\% lower than GPT-4's rate using the same prompt strategy. While most LLMs demonstrate comparable performances on easy tasks, their success rates vary significantly more on medium and hard tasks. All LLMs, excluding GPT-4, can solve at most half of the medium tasks and only 15\% of the hard tasks within five attempts, which is substantially lower than the success rate of GPT-4 with the optimal prompt strategy.

\subsection*{Comparing coding performance of LLMs and human programmers}

We then compared the coding performance of LLMs and human participants in LeetCode and GeeksforGeeks contests. In these contests, participants are asked to solve several programming tasks in a given period of time, and are ranked based on the tasks successfully solved, the number of attempts taken to solve the tasks, and the total amount of time spent. After the contest ends, the statistics and rankings of participants are published and can be easily queried. Plagiarism and cheating are strictly prohibited and will disqualify a participant if detected. Thus, these contests provide a way to rigorously benchmark the performance of LLMs with human programmers. Supplementary Table 1-2 provide a comprehensive compilation of the prompts used, the programming code generated by the LLMs under different prompt strategies, and the evaluation results in both LeetCode and GeeksforGeeks contests.

LLMs participated in the contests in a mock manner, with us serving as the interface between the LLMs and the contest websites. The best-performing strategy, the Python3 feedback CI prompt, was used for GPT-4. For other LLMs, the prompt strategies were the same as those in Figure 2. Each LLM was given a maximum of five attempts for each task. It is important to note that the contests do not impose a restriction on the number of attempts allowed. Thus, the performance of LLMs could potentially be further improved with more attempts. After each LLM completed the contest, we recorded its performance and followed the same competition rules to calculate the scores and obtain the ranking for each LLM (Methods).

Figure 3a and Supplementary Table 3 show the percentile rankings of all LLMs. GPT-4 significantly outperforms other LLMs, consistent with the previous results shown in Figure 2. For most other LLMs, the percentile ranking is around 60\% for most coding contests, indicating that their coding capabilities are comparable to those of average human programmers. Figure 3b displays GPT-4's percentile rankings, and the total number of participants for each coding contest. GPT-4 demonstrates highly consistent performance in both LeetCode and GeeksforGeeks contests, indicating that the performance benchmark is reproducible. In 18 out of 26 contests (69\%), GPT-4 ranks above the 90th percentile. In 21 out of 26 contests (81\%), GPT-4 ranks above the 85th percentile. In three LeetCode contests with more than 18,000 participants, GPT-4 ranks within the top 100. These results suggest that GPT-4 possesses superior programming abilities compared to most human programmers participating in these contests, yet it still falls short of surpassing the most elite human programmers in most situations.

Figure 3c further compares the success rates of LLMs across programming tasks categorized by the proportion of human programmers who can successfully solve each task. More than 40\% of human programmers fail the easiest task. The success rates of LLMs increase for tasks solvable by a larger percentage of human programmers. For tasks that more than 20\% of human programmers can solve, GPT-4's success rates exceed 90\%. Even for tasks that less than 10\% of human programmers can solve, GPT-4 still achieves a success rate of 33\%. These findings suggest that GPT-4 can serve as a reliable assistant, enhancing the coding performance of most human programmers.

\subsection*{LLMs' ability to learn from error messages in programming code}

Figure 2 shows that five-attempt performances are significantly better than one-attempt performances in most cases, suggesting that running additional attempts is needed to boost the success rate of programming tasks. However, the performance gain is different for different LLMs and prompt strategies. For GPT-4, prompt strategies that rely on feedback messages benefit more from additional attempts. For the feedback CI prompt strategy, the five-attempt success rates are 0.25 and 0.18 points higher than the one-attempt success rates for medium and hard LeetCode tasks, respectively. These differences are reduced to 0.14 and 0.08, respectively, for the repeated prompt strategy.

To better understand this behavior of LLMs, Figure 4a shows the salvage rate, defined as the success rate of subsequent attempts for programming tasks that failed the initial attempt, comparing GPT-4 with different prompt strategies. For the repeated prompt and multiway prompt, running a second attempt salvages a considerable number of programming tasks that fail the first attempt, but running more than two attempts has a marginal effect on further increasing the success rate. In comparison, both the feedback prompt and the feedback CI prompt continue to benefit from additional attempts, eventually salvaging over 60\% of tasks with easy and medium difficulties that failed the first attempt. The feedback CI prompt has the best overall salvage rate for hard tasks compared to other prompt strategies and is able to salvage over 20\% of hard tasks that fail the first attempt. These results suggest that GPT-4 is able to learn from the error messages and can iteratively fix its own coding errors with repeated attempts. In practical terms, running five attempts is generally sufficient, as the curves for both the feedback CI prompt and the feedback prompt tend to plateau beyond this point.

Figure 4b further compares the salvage rates of different LLMs. GPT-4 exhibits the strongest performance in learning from previous errors, while Claude 2 and GPT-3.5 also show considerable abilities, particularly in easy tasks. In contrast, other LLMs do not demonstrate a clear ability to learn from previous errors.

\subsection*{GPT-4's ability to translate across programming languages}

Figure 2 also shows that the Python3 translation strategy outperforms the feedback prompt for Java, JavaScript, and C++, suggesting an alternative approach to improve the success rate of programming languages that cannot access GPT-4's CI function (Figure 5a). This is dependent on both the superior performance of the feedback CI prompt and GPT-4's high accuracy in translating across programming languages.

Figure 5b and Supplementary Table 4 show the success rate of different programming languages translated from Python3 by GPT-4. The original programming task and the Python3 code output from a previous GPT-4 query for solving the task were given to GPT-4 when GPT-4 was asked to perform the translation. When the original Python3 code is correct, GPT-4 accurately translates it in almost all tasks. The success rate is almost identical across different target languages. Surprisingly, GPT-4 is still able to generate code that correctly solves the programming task even when the original Python3 code is incorrect in some medium tasks. These results suggest that GPT-4 can serve as a reliable tool to translate code across programming languages in most cases.


\subsection*{Computational efficiency of code generated by GPT-4}

Finally, we found that the running time and memory usage of code generated by GPT-4 are comparable to those of human programmers when GPT-4 was not specifically instructed to optimize computational efficiency (Figure 5c, Supplementary Table 5). We then asked GPT-4 to optimize the computational efficiency of the code it had previously generated (Methods). Both running time and memory usage improve slightly in this case (Figure 5c). These results suggest that the code generated by GPT-4 is not specifically optimized for computational efficiency, and indicate that GPT-4 has a limited capability in improving the computational efficiency of existing codes.



\section*{Discussion}

In this study, we found that GPT-4 is the best-performing LLM in generating programming code. GPT-4's performance is substantially influenced by various prompt strategies. The best prompt strategy utilizes GPT-4's code interpreter function to test the validity of example test cases and benefits from the error messages from previous unsuccessful attempts. In LeetCode and GeeksforGeeks contests, GPT-4 exhibits stronger coding performance than most human programmers, while other LLMs perform similarly to average human programmers. GPT-4 demonstrates reliable coding performance across various programming languages and shows a strong ability to translate code between different programming languages.

These results suggest that GPT-4 may serve as a reliable assistant in generating programming code for practical applications. GPT-4 may empower individuals with little or no programming expertise to solve programming tasks of easy or medium difficulty. For those who have advanced programming expertise, GPT-4 may share the workload, allowing human programmers to focus on more challenging tasks. Thus, GPT-4 introduces a new paradigm in human-AI collaboration, paving the way for a more efficient and inclusive process in generating programming code.

It is worth noting that generating programming code is only a subarea of software engineering. The ability of LLMs to manage projects, design system architectures, create user interfaces, test and maintain code, and perform many other tasks related to software engineering has not been evaluated in this study. In addition, LLMs relies on well-documented descriptions of programming tasks and comprehensive sets of test cases for validating the generated programming code. This reliance may necessitate additional effort and expertise in writing prompts and establishing a software testing infrastructure.

\section*{Methods}

\subsection*{Prompt construction}

Each LeetCode programming task was copied directly from the LeetCode website. The task includes the description of the problem (Supplementary Figure 3a), multiple example cases with inputs, desired outputs, and explanations (Supplementary Figure 3b), numerical constraints of parameters (Supplementary Figure 3c), and a code template consisting of several lines of code (Supplementary Figure 3d). The following sentence was added before the code template: ``Code starts with:''. Each GeeksforGeeks programming task was manually transcribed from the GeeksforGeeks website, with the information organized in the same manner as the LeetCode tasks.

To construct the prompt for repeated prompt strategy with example test cases, the following sentence was added to the top of the full LeetCode programming task: ``write Python3 code for the following question: ''.

For the repeated prompt strategy without example test cases, the same prompts as the repeated prompt strategy with example test cases were used, except the example test cases were removed.

For the multiway prompt strategy, the following sentence was added to the top of the full LeetCode programming task: ``write Python3 code for the following question using five different ways: ''.

For the feedback prompt strategy, the first attempt uses the same prompt as the repeated prompt strategy, where ``Python3'' is replaced with ``Java'', ``Javascript'', or ``C++'' for other programming languages. For subsequent prompts, error messages (Supplementary Figure 3e) were directly copied and pasted from LeetCode to GPT-4.

For the feedback CI prompt strategy, the initial prompt is identical to that of the feedback prompt, except that the sentence ``Use a code interpreter to check the following examples:'' is added above the example test cases. In subsequent attempts, if the error message from the previous attempt pertains to incorrect test cases, the new prompt will start with 'Use a code interpreter to check the following examples:', followed by the example test cases from the original programming task and all incorrect test cases from previous failed attempts. If the error message is related to other issues, the error message is directly copied and pasted from LeetCode to GPT-4.

For translating Python3 programming code into another programming language, the prompt is composed of the following components in order: ``Convert the following Python3 code into Java''; ``The Java Code starts with:''; a Java code template copied from LeetCode; ``The Python3 code:''; the Python3 code generated previously by the GPT-4 feedback CI prompt strategy for solving the programming task; ``The original task was:''; and the prompt used in the Python3 repeated prompt strategy. ``Java'' is replaced with ``Javascript'' or ``C++'' for other programming languages.

To optimize the memory usage and running time of Python3 code, the following sentence was added above the Python3 code generated previously by the GPT-4 feedback CI prompt strategy: ``Improve the memory usage and running time of the following Python3 code''.

\subsection*{Query of GPT-4}

For GPT-4, all programming tasks in this study were tested using the ``gpt-4-0613'' model. GPT-4 queries were performed using the OpenAI Playground online web portal (\url{https://platform.openai.com/playground}).

For the multiway prompt strategy, GPT-4 was queried once. For all other prompt strategies, GPT-4 was queried up to five times.

The repeated prompt strategy and the feedback prompt strategy begin with the same initial prompt. For each task, we first used the feedback prompt strategy. The output from the first attempt of the feedback prompt strategy was then directly used as the first attempt's output for the repeated prompt strategy, rather than starting the latter from scratch. The repeated prompt strategy was then allowed up to four additional attempts. Therefore, the one-attempt performances of the feedback and repeated prompt strategies are identical. This approach ensures that the comparison between these two strategies is not influenced by the randomness in the results of the first attempts.

\subsection*{Query of other LLMs}

GPT-3.5 queries were performed using the ``gpt-3.5-turbo-0613'' model with the OpenAI Playground online web portal (\url{https://platform.openai.com/playground}).

Gemini Ultra 1.0 and Gemini Pro were accessed with the Gemini website (\url{https://gemini.google.com/}) from February 10 2024 to February 19 2024.

Claude 2 (Claude-2-100k), Llama 2 (Llama-2-70b), and Code Llama (Code-Llama-34b) were accessed with the Quora Poe app (\url{https://poe.com/}) from January 5 2024 to February 1 2024.

\subsection*{Scoring and ranking of LeetCode contests}

LeetCode assigns a numerical score to each of the four programming tasks in a LeetCode contest. The overall contest score is the sum of the scores for tasks successfully solved. To calculate the total time an LLM spent on a contest, we consider the time for a complete LLM attempt as one minute, although in real practice, it often takes less than one minute. A complete LLM attempt includes preparing the prompt by copying task information from LeetCode, querying the LLM, transferring the programming code from the LLM output to LeetCode, and running the evaluation on LeetCode. Per LeetCode contest rules, each failed attempt incurs a penalty time of 5 minutes. Therefore, if an LLM solves the task at the $k$th attempt ($k \leq 5$), the time the LLM spent on that task is $k+(k-1)*5$ minutes. The overall time for a contest is the sum of the time spent on tasks successfully solved.

LeetCode ranks participants by their overall scores in descending order. For participants with tied scores, LeetCode ranks participants by their overall time spent in increasing order. We manually queried the ranking list of human participants by LeetCode and identified the ranking of an LLM. If an LLM achieves an overall score of 0, its ranking is assigned as the mean ranking of all human participants who also scored 0.

\subsection*{Scoring and ranking of GeeksforGeeks contests}

GeeksforGeeks assigns a numerical score to each of the three programming tasks in a GeeksforGeeks contest. The overall contest score is the sum of the scores for problems successfully solved. Each contest problem's score incurs a 5\% penalty for each wrong submission. For example, the score for a problem originally worth 40 points will be reduced to 38 points if there is one wrong submission.

GeeksforGeeks ranks participants according to their overall scores in descending order. In cases of tied scores, participants are further ranked based on the date and time of their last correct submission, with earlier submissions receiving higher rankings. However, GeeksforGeeks does not account for the time taken to complete the contest. Since there is no penalty for completion time, we determined an LLM's rank to be that of the best-ranked human participant with the same overall score, assuming that all LLMs generate programming code faster than any human participant. GeeksforGeeks does not maintain records for participants with an overall score of 0. If an LLM achieves an overall score of 0, its ranking is determined by adding one to the total number of human participants who have positive overall scores.

\section*{Acknowledgments}
Z.J. is supported by the Whitehead Scholars Program at Duke University School of Medicine.  W.H. is supported by the General Fund at Columbia University Department of Biostatistics. 

\section*{Author contributions}
Z.J. and  W.H. both conceived the study, conducted the analysis, and wrote the manuscript.

\section*{Competing interests}
All authors declare no competing interests.

\section*{Data availability}
All data generated or analyzed during this study are included in the supplementary tables.

\clearpage

\begin{figure}[t]
    \centering
    \includegraphics[width=\linewidth]{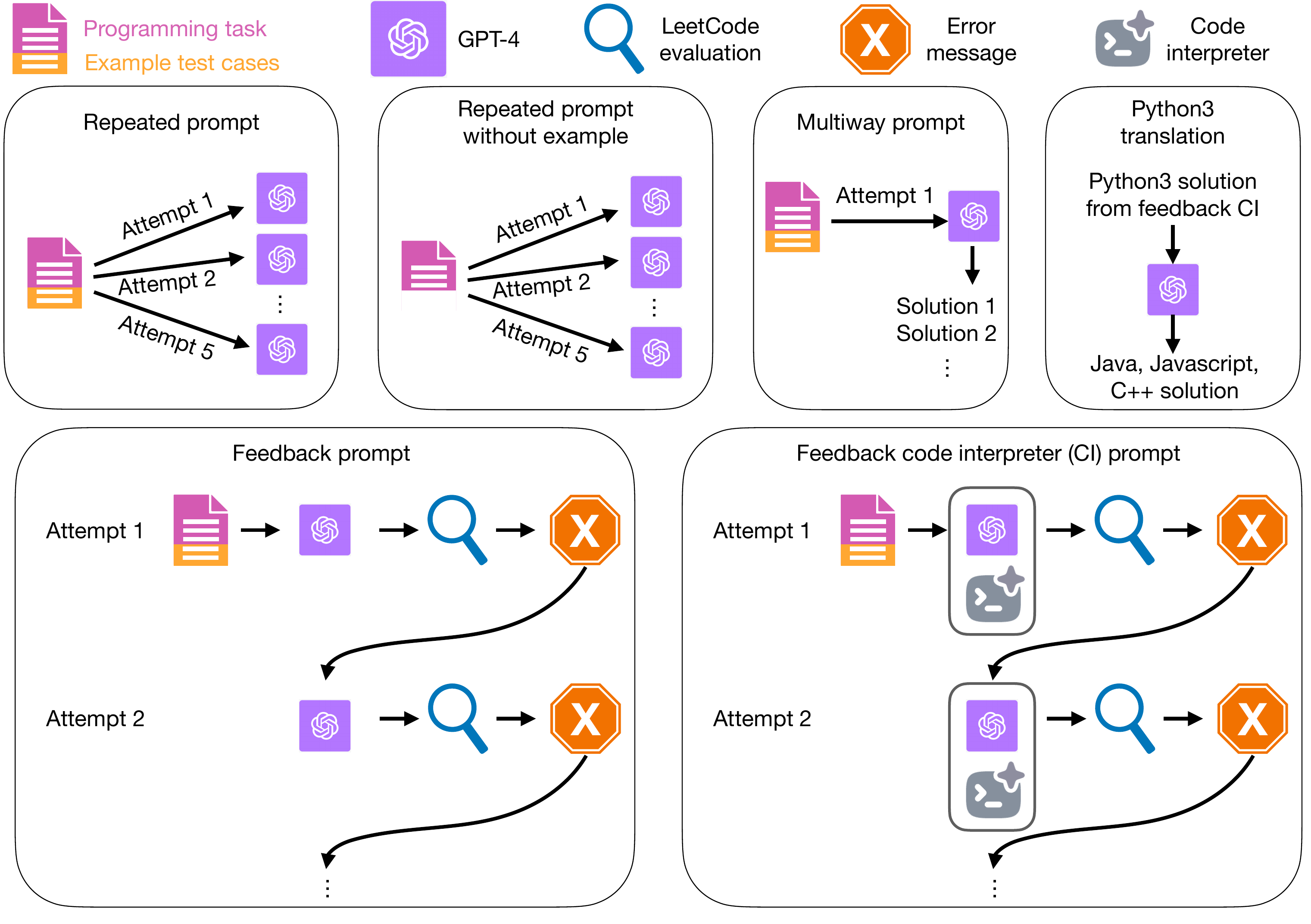}
    \caption{A schematic illustrating the prompt strategies that were developed and assessed in this study.}
    \label{fig:intro}
    \vspace{-5mm}
    \end{figure}

\clearpage

\begin{figure}[t]
    \centering
    \includegraphics[width=\linewidth]{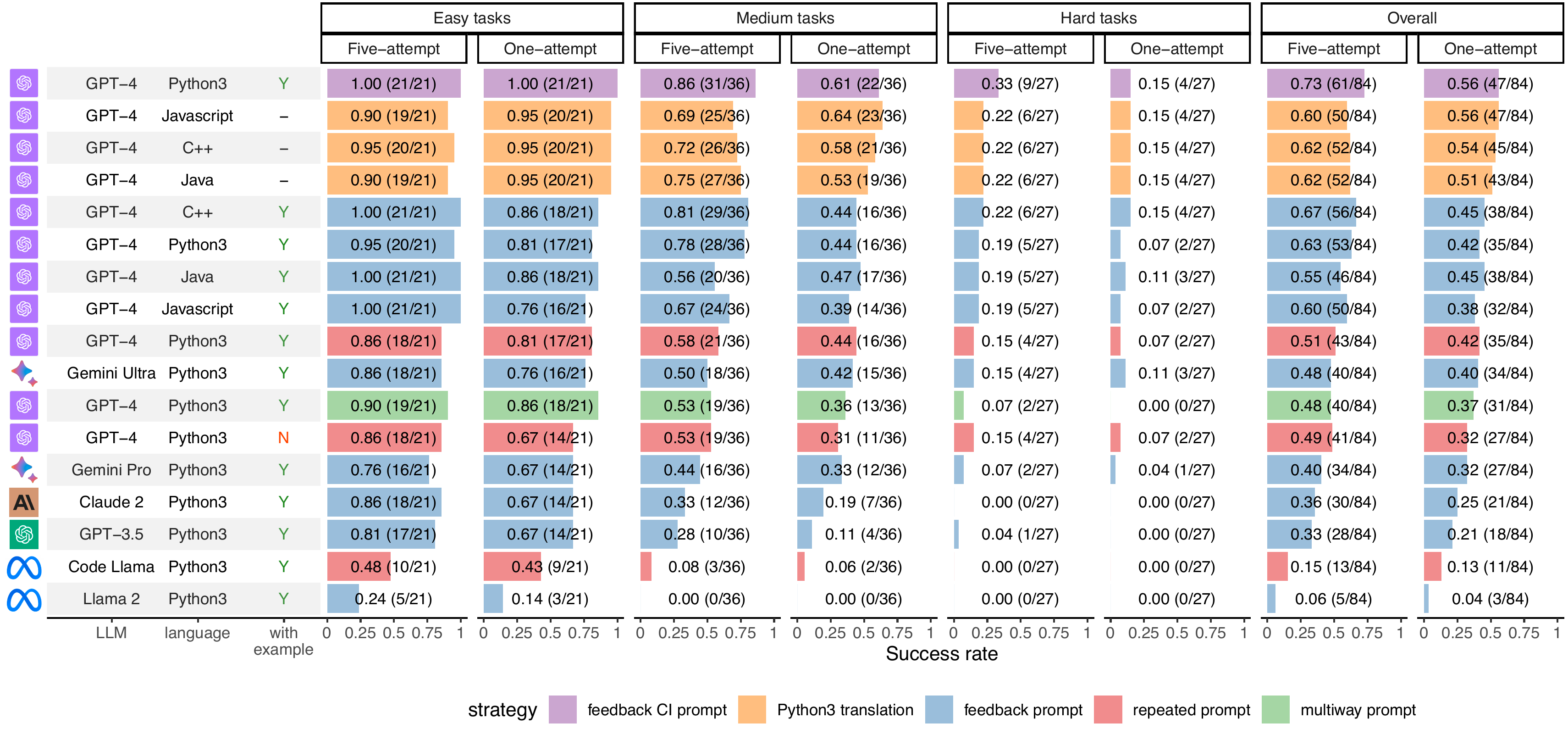}
    \caption{Five-attempt and one-attempt success rates for tasks with varying difficulty levels. Each row represents a different prompt strategy and a different LLM. The text within each bar displays the success rate, the number of programming tasks solved, and the total number of tasks. Rows are ordered according to the average success rates for both five attempts and one attempt across all tasks.}
    \label{fig:perf}
    \vspace{-5mm}
    \end{figure}

\clearpage

\begin{figure}[t]
    \centering
    \includegraphics[width=\linewidth]{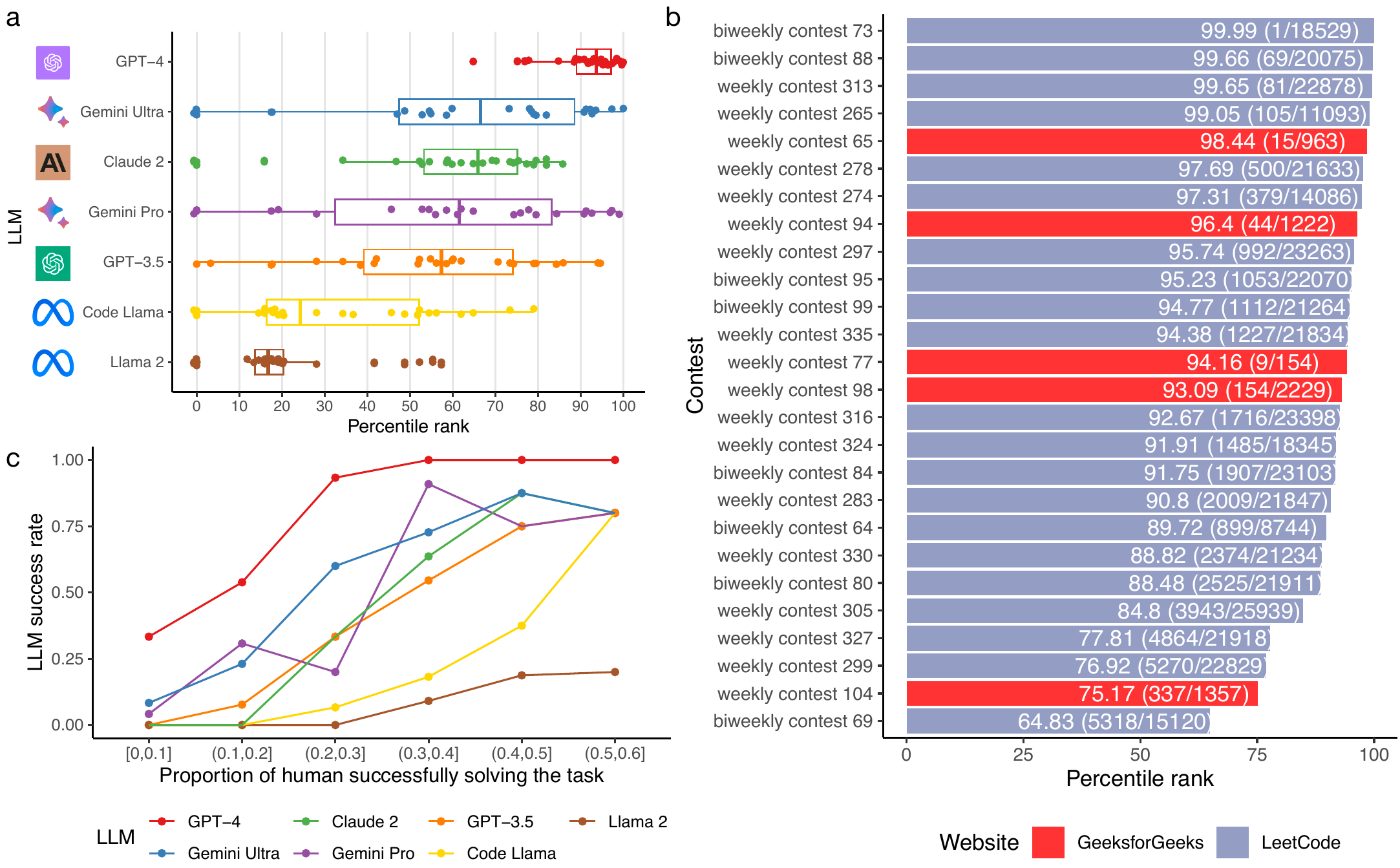}
    \caption{Comparing coding performances of LLMs and human programmers. \textbf{a}, Percentile rank (x-axis) of LLMs (y-axis) for LeetCode and GeeksforGeeks contests. Each dot is a coding contest. \textbf{b}, Percentile rank of GPT-4 for LeetCode and GeeksforGeeks contests, colored differently. Each row represents a contest. Texts in the bar show the percentile rank of GPT-4, the rank of GPT-4, and total number of participants. \textbf{c}, LLM success rates (y-axis) for LeetCode programming tasks categorized by proportion of human successfully solving the task (x-axis). LLMs are represented by different colors.}
    \label{fig:translate}
    \vspace{-5mm}
    \end{figure}

\clearpage

\begin{figure}[t]
    \centering
    \includegraphics[width=\linewidth]{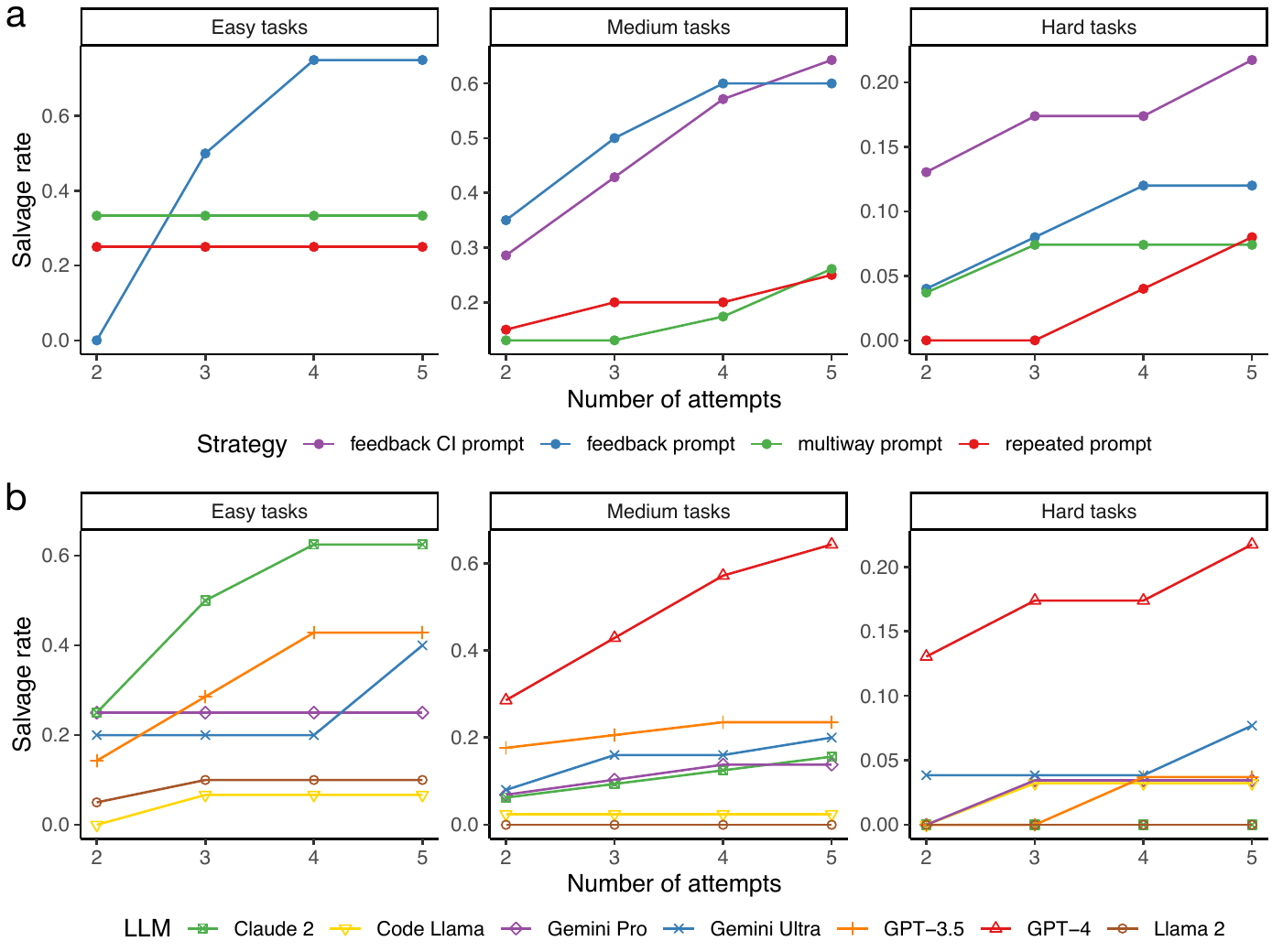}
    \caption{The salvage rates (y-axis) increase with the number of attempts (x-axis) for GPT-4 using different prompt strategies (\textbf{a}) and for different LLMs (\textbf{b}). In \textbf{b}, the feedback CI prompt strategy was used for GPT-4, and the prompt strategies for other LLMs were the same as those in Figure 2.}
    \label{fig:numshot}
    \vspace{-5mm}
    \end{figure}

\clearpage

\begin{figure}[t]
    \centering
    \includegraphics[width=\linewidth]{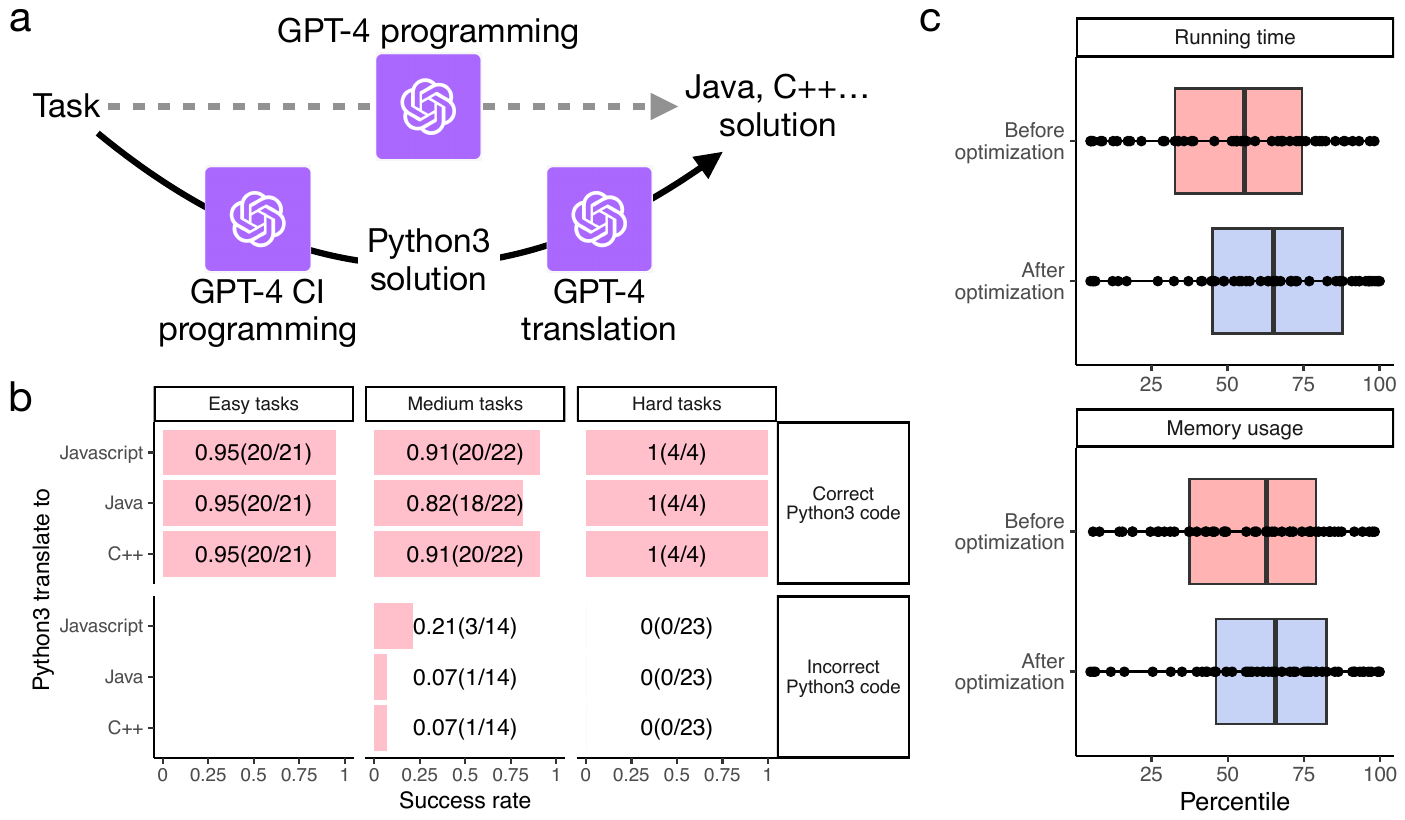}
    \caption{Evaluation of GPT-4's abilities in translating across programming languages and the computational efficiency of the generated code. \textbf{a}, A schematic illustrates a proposed strategy to enhance the success rate of non-Python3 programming languages using translation. \textbf{b}, The success rate of translating Python3 code, generated from one-attempt feedback CI prompts, to other programming languages for tasks of varying difficulty levels. The top and bottom panels display results for when the original Python3 code is correct or incorrect, respectively. Since Python3 code generated by GPT-4 is correct for all easy tasks, translation is not evaluated for easy tasks when the Python3 code is incorrect. \textbf{c}, Running time and memory usage percentile for GPT-4 generated code compared to human programmers before and after optimization by GPT-4. A higher percentile represents lower running time and memory usage, indicating better computational efficiency.}
    \label{fig:numshot}
    \vspace{-5mm}
    \end{figure}

\clearpage

\bibliography{bibfile.bib}   
\end{document}